\documentclass[twocolumn,showpacs,amsmath,amssymb,pre,floatfix]{revtex4}
\usepackage[pdftex]{epsfig,graphicx}% Include figure files 
\usepackage{dcolumn}% Align table columns on decimal point
\usepackage{bm,ams}% bold math
\usepackage{rjwmath}
\usepackage{ifpdf}

\begin{document}

\title{Enhancement of electron spin lifetime in GaAs crystals: the benefits of dichotomous noise}

\author{Stefano Spezia\footnote{Email: stefano.spezia@gmail.com}, Dominique Persano Adorno, Nicola Pizzolato, Bernardo Spagnolo}
 \affiliation{Dipartimento di Fisica e Chimica, \\
Universit\`a di Palermo and CNISM-INFM, \\
Viale delle Scienze, edificio 18, I-90128 Palermo, Italy}

\begin{abstract}
The electron spin relaxation process in n-type GaAs crystals driven by a fluctuating electric field is investigated. Two different sources of fluctuations are considered: (i) a symmetric dichotomous noise  and (ii) a Gaussian correlated noise. Monte Carlo numerical simulations show, in both cases, an enhancement of the spin relaxation time by increasing the amplitude of the external noise. Moreover, we find that the electron spin lifetime versus the noise correlation time: (i) increases up to a plateau in the case of dichotomous random fluctuations, and (ii) shows a nonmonotonic behaviour with a maximum in the case of bulks subjected to a Gaussian correlated noise.
\end{abstract}

\maketitle

\section{Introduction}

An emergent spin-based electronics technology, where the information is carried by the electron spin, offers the way to enhance the functionalities of the current electronics by controlling spin by electric currents or gate voltages~\cite{Yu2002,Zutic2004,Wolf2008,Insight2012,Salahuddin2013,Awschalom2013}.

The major challenge of semiconductor spintronics is the development of spin-based devices in which the binary information is encoded in the two spin-states "up" and "down". These states are transferred and manipulated by applying electric and/or magnetic fields, and lastly detected~\cite{Awschalom2007}. However, a disadvantage of the utilization of spin degree of freedom is that the spin of conduction electrons decays over time during the transport because of the combined effect of spin-orbit coupling and momentum scattering. Thus, the spin relaxation time could be inadequately short to allow the completion of the necessary spin manipulations. Therefore, a full investigation of spin relaxation dynamics becomes an essential point in the design of spintronic devices~\cite{Zutic2004,Flatte2009}.\\
\indent  Recently, noise-induced complex phenomena in nonlinear systems have increasingly been investigated, with a focus on cooperative effects between the noise and the intrinsic interactions of the 
system~\cite{Ghosh2008,Li2010,Ghosh2012,Nakada2012,Wang2013,Sen2013}.
In particular, in the last decade, great interest has been oriented towards the possible positive effects of noise on nonlinear systems. Previous theoretical studies have revealed that, under specific conditions, the addition of external noise sources to intrinsically noisy systems may induce an enhancement of the dynamical stability of the system, resulting in a less noisy response~\cite{Landa2000,Vilar2001,Walton2004,Seol2004,Dubkov2004,Fiasconaro2009}. 
In Refs.~\cite{Atxitia2009,Bose2010}, a way to improve the ultra-fast magnetization dynamics of magnetic spin systems by including random fields has been discussed. Noise enhanced stability induced by a superimposed source of noise in the electron transport in GaAs crystals, subjected to periodic electric fields, has been 
found~\cite{Persano2009,Persano2012}. In semiconductor quantum wells and wires, Glazov et al. have demonstrated that the randomness in spin-orbit coupling is inevitable and can be attributed both to the electron-electron dynamic collisions and the static fluctuations in the density of dopant ions~\cite{Glazov2010,Glazov2011}. Furthermore, they pointed out the possibility of using fluctuating random Rashba spin-orbit interaction for the generation of spin currents~\cite{Dugaev2012}. Monte Carlo simulations have also shown that random spatial variation of the Rashba electric field along the length of a quantum
wire makes the spatial spin relaxation characteristics random, non-monotonic and chaotic~\cite{Agnihotri2012}.\\
\indent In a previous study of the electron spin relaxation process in GaAs bulks, at nitrogen temperature, we have shown that a random contribution added to the static electric field can affect the spin decoherence length ~\cite{Spezia2012}. In particular, it has been found that the effect on spin depolarization length is maximum for values of the noise correlation time comparable with the characteristic time of spin relaxation process, and that, depending on the amplitude of the applied electric field, the external fluctuations can have opposite effects~\cite{Spezia2012}.\\
\indent Aim of the present work is to study the effects on the electron spin relaxation process of a different kind of external noise. In particular, here we focus on the influence of a non-Gaussian fluctuating contribution to the driving electric field: a random telegraph noise source. By superimposing this noise source to the intrinsic one, it is possible to tune the dynamic response of the system. The electron dynamics is simulated by a semiclassical Monte Carlo approach, which takes into account all the possible scattering events of the hot electrons in the medium~\cite{Persano2000,Persano2010} and includes the precession equation of the spin polarization vector~\cite{Saikin2006,Spezia2010_2}. Starting from an initial spin polarization $\vect{S}(0) = 1$, with all spins in the same direction, we calculate the relaxation time as a function of the characteristic parameters of the external noise source. For intense electric fields, we find that both in the presence of dichotomous noise and a Gaussian correlated noise, the spin relaxation time can be enhanced within a wide range of noise correlation times.

\section{Electron spin relaxation model}

The spin-orbit interaction couples the spin of conduction electrons to the electron momentum, which is randomized by scattering with phonons, impurities and other carriers. The spin-orbit coupling gives rise to a spin precession, while momentum scattering makes this precession randomly fluctuating, both in magnitude and orientation~\cite{Dyakonov2006}.\\
\indent For delocalized electrons and under nondegenerate regime, the D'yakonov-Perel (DP) mechanism \cite{Dyakonov2006} is the only relevant relaxation process in n-type III-V semiconductors~\cite{Litvi2010}. In a semiclassical formalism, the term of the single electron Hamiltonian which accounts for the spin-orbit interaction can be written as $H_{SO} = \frac{\hbar}{2}\vect{\sigma}\cdot\vect{\Omega}$. It represents the energy of electron spins precessing around an
effective magnetic field ($\vect{B}=\hbar\vect{\Omega}/\mu_Bg$) with angular frequency $\vect{\Omega}$, which depends on the orientation of the electron momentum vector with respect to the crystal axes ($\mu_B$ is the Bohr magneton and $g$ is the electron spin g-factor). Near the bottom of each valley, the precession vector can be written as
\begin{equation}
\vect{\Omega}_{\Gamma}=\frac{\beta_{\Gamma}}{\hbar}
[k_{x}(k_{y}^{2}-k_{z}^{2})\hat{x}+k_{y}
(k_{z}^{2}-k_{x}^{2})\hat{y}+k_{z}(k_{x}^{2}-k_{y}^{2})\hat{z}]
\label{effectivefieldgammavalley}
\end{equation}
in the $\Gamma$-valley~\cite{Dresselhaus55} and
\begin{equation}
\vect{\Omega}_{L}=\frac{\beta_{L}}{\sqrt{3}}
[(k_{y}-k_{z})\hat{x}+(k_{z}-k_{x})\hat{y}+(k_{x}-k_{y})\hat{z}]
\label{effectivefieldLvalley}
\end{equation}
in the L-valleys located along the [111] direction in the crystallographic axes~\cite{Saikin2006}. In equations
(\ref{effectivefieldgammavalley})-(\ref{effectivefieldLvalley}), $k_{i}$ ($i=x,y,z$) are the components of the electron wave vector,
$\beta_{\Gamma}$ and $\beta_{L}$ are the spin-orbit coupling coefficients. Here, we assume $\beta_{\Gamma}$= $23.9$ eV $\cdot${\AA}$^{3}$, as used in Ref.~\cite{Tong2012} and $\beta_{L}$=$0.26$ eV $/${\AA}$\cdot2/\hbar$, as recently theoretically estimated in Ref.~\cite{Fu2008}.\\
\indent Since the quantum-mechanical description of the electron spin evolution is equivalent to that of a classical
momentum $\vect{S}$ experiencing the magnetic field $\vect{B}$, we describe the spin dynamics by the classical equation of precession motion $\frac{d\vect{S}}{dt}=\vect{\Omega}\times\vect{S}$. The DP mechanism acts between two scattering events and reorients the direction of the precession axis and the effective magnetic field $\vect{B}$ randomly and in a trajectory-dependent way. This effect leads the spin precession frequencies $\vect{\Omega}$ and their directions to vary in an inhomogeneous way within the electron spin ensemble. This spatial variation, called {\it inhomogeneous broadening}, is quantified by the average squared precession frequency $\langle\mid\vect{\Omega}(\vect{k})\mid^2\rangle$~\cite{Slichter}. This quantity, together with the correlation time of the random angular diffusion of spin precession vector $\tau_\Omega$, are the relevant variables in the D'yakonov-Perel's formula \cite{Dyakonov2006}
\begin{equation}
\tau=\frac{1}{\langle\mid\vect{\Omega}(\vect{k})\mid^2\rangle\tau_\Omega}~.
\label{DPformulae}
\end{equation}
Here, by following Matthiessen's rule, $1/\tau_\Omega=1/\tau_p+1/\tau^{'}_p$, with $\tau_p$ the momentum relaxation time and $\tau^{'}_p$, the momentum redistribution time, related to the
electron-electron interaction mechanism. This distinction is necessary because, although electron-electron scattering contributes to momentum redistribution, it does not directly lead to momentum relaxation~\cite{Glazov2002}. The spin relaxation time $\tau$ results inversely proportional to both the correlation time of the fluctuating spin precession vector
$\tau_\Omega$ and the inhomogeneous broadening $\langle\mid\vect{\Omega}(\vect{k})\mid^2\rangle$.

\section{Noise modelling and numerical approach}
 
In our simulations the semiconductor bulk is driven by a fluctuating electric field $F(t)=F_0+\eta(t)$, where $F_0$ is the
amplitude of the deterministic part and $\eta(t)$ is the random contribution due to an external noise source. Here we consider two different kinds of noise source: a dichotomous Markov (DM) noise and a Gaussian correlated (GC) noise.\\
\indent The DM noise is generated by a random process taking only discrete values and stochastically switching between these values. We consider a symmetric dichotomous Markov stochastic process with only two values \cite{Bena2006,Barik2006}
\begin{equation}
\eta(t)\in\{-\Delta,\Delta\}.
\label{dichomotous1}
\end{equation}
Thus, we have a zero mean
\begin{equation}
\langle\eta(t)\rangle=0,
\label{dichomotous2}
\end{equation}
and correlation function
\begin{equation}
\langle\eta(t)\eta(t')\rangle=\Delta^2\exp{\left(-\frac{\mid t-t'\mid}{\tau_D}\right)},
\label{dichomotous3}
\end{equation}
where $\tau_D$ is the correlation time of the noise and it is related to the inverse of the mean frequency of transition from $\pm\Delta$ to $\mp\Delta$, respectively~\cite{Bena2006,Barik2006}. In our runs, we choose $\eta(0)=X$ as initial condition, where $X$ is a random variable which takes the values $-\Delta$ and $\Delta$ with equal probability ($p=1/2$). We consider only fluctuations of equal height, in such a way that this external noise can be easily generated in practical systems and tuning effects can be more controllable. A dichotomous Markovian noise can be realized, for example, by means of a cheap and simple, home-made, noise generator, based on the generation of a pseudo-random sequence by a linear-feedback shift register SR2, as extensively described in Ref.~\cite{Fronzoni1989}.
\\
\begin{figure*}
\includegraphics[scale=0.52]{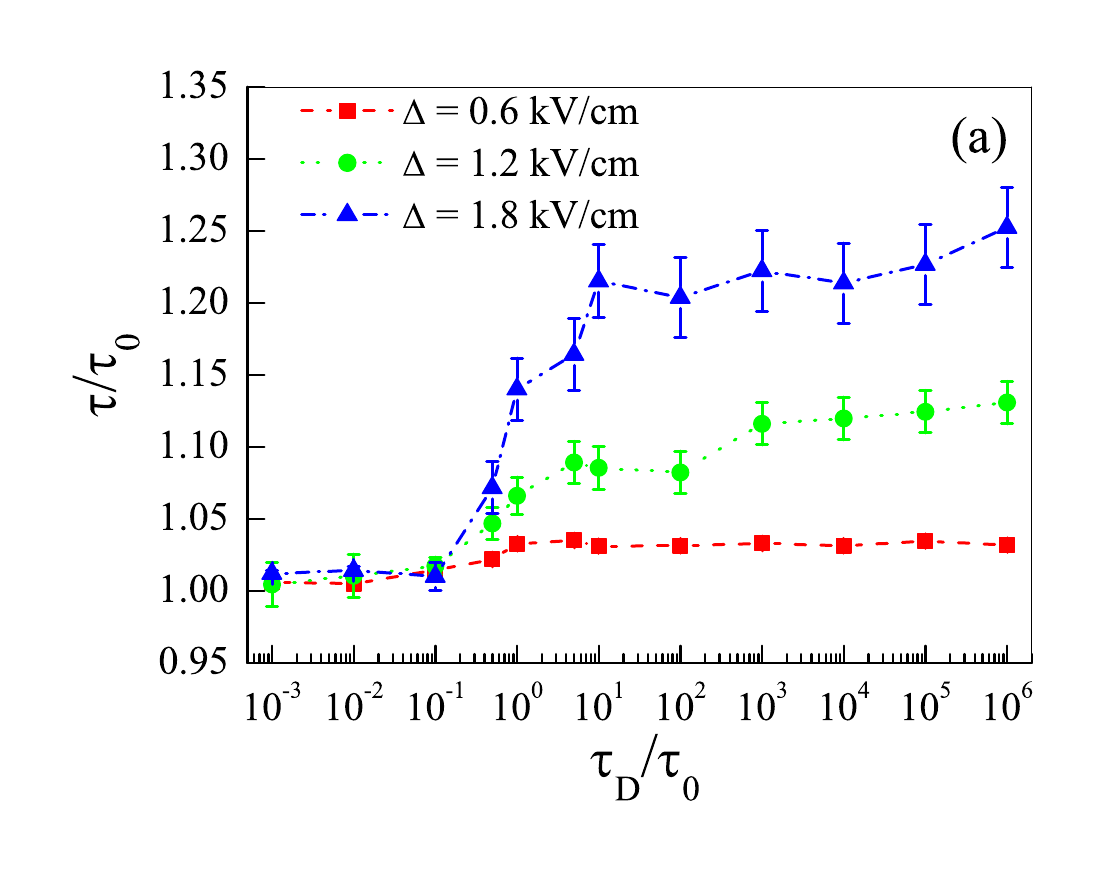}
\includegraphics[scale=0.52]{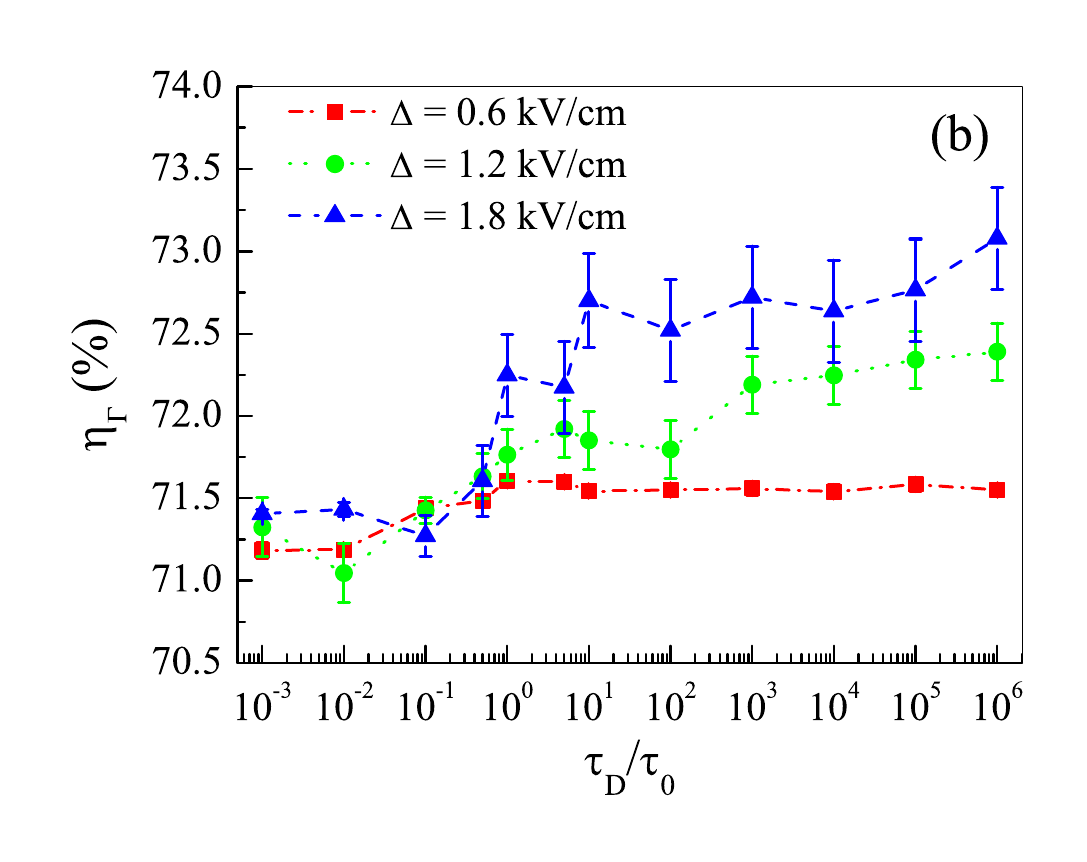}
\includegraphics[scale=0.52]{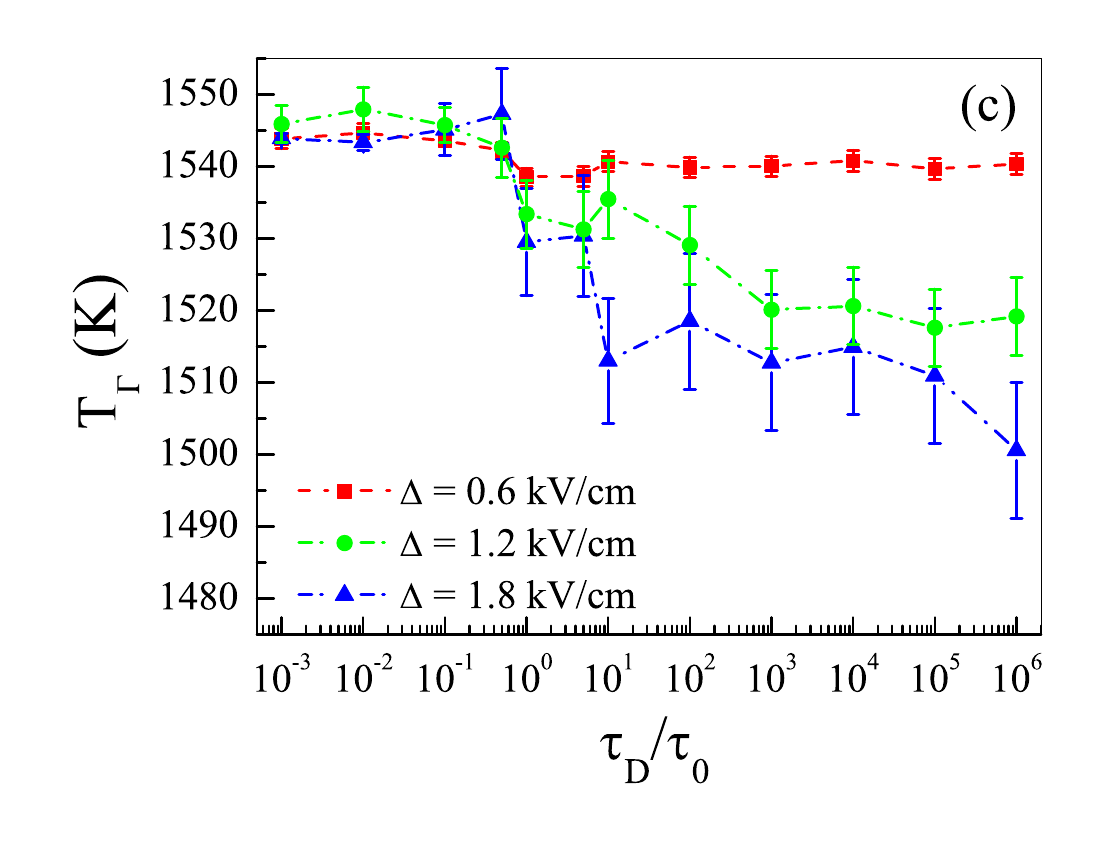}
\caption{Dichotomous noise results: (a) Normalized electron spin relaxation time $\tau/\tau_0$, (b) electron occupation percentage $\eta_\Gamma$ and (c) hot-electron temperature $T_\Gamma$ in $\Gamma$-valley, as a function of the normalized noise correlation time $\tau_D/\tau_0$, for three different values of noise amplitude, namely $\Delta = 0.6, 1.2, 1.8$ kV/cm. The values of the other parameters are: $n=10^{16}$ cm$^{-3}$, $T_L$=$300$ K, $F_0=6$ kV/cm and $\tau_0=1.40$ ps.}
\label{fig.1}
\end{figure*}
\indent Gaussian correlated noise is modelled as an Ornstein-Uhlenbeck (OU) process, which obeys the following stochastic
differential equation
\begin{equation}
\textrm{d}\eta(t)=
-\frac{\eta(t)}{\tau_c}\textrm{d}t + \sqrt{\frac{2D}{\tau_c}}\textrm{d}W(t)
\label{Ornstein}
\end{equation}
where $\tau_c$ and $D$ are the correlation time and the intensity of the noise, respectively. The autocorrelation function of the OU process is $\langle\eta(t)\eta(t')\rangle=D\exp(-|t-t'|/\tau_c)$, and W(t) is the Wiener process with the usual statistical properties $\textrm{d}W(t) = 0$ and $\langle \textrm{d}W(t)\textrm{d}W(t')\rangle = \delta(t-t')$.
A correlated Gaussian noise can be easily realized in a practical system. It, for example,  could be generated by a RC circuit driven by a source of Gaussian white noise, with correlation time $\tau_c$=(RC)$^{-1}$, as discussed in Ref.~\cite{Spezia2012}.\\
\indent In this work, the considered conduction bands of GaAs are the $\Gamma$-valley and four equivalent L-valleys. We do not include the three equivalent X-valleys because, in the investigated range of values of the driving electric field, their occupation remains always lower than 1$\%$.  The Monte Carlo procedure used to simulate the electron transport takes into account the effects of intravalley and intervalley scattering of electrons in multiple energy valleys, and those caused by the nonparabolicity of the band structure. Electron scatterings due to ionized impurities, acoustic, piezoelectric and polar optical phonons in each valley as well as all intervalley transitions between equivalent and nonequivalent valleys are accounted for~\cite{Persano2010}. Moreover, we also include the carrier-carrier interaction by using the screened Coulomb potential and the Born approximation. In particular, the electron-electron scattering is
treated as an interaction between two particles, by using the
Peschke's approach~\cite{Peschke94}, as refined by Mo\v{s}ko and Mo\v{s}kov\'{a} in order to take into account the electron-electron scattering rate valid for spin-polarized gas~\cite{Mosko91}. The complete set of n-type GaAs parameters used in our calculations is listed in Table I of Ref.~\cite{Persano2000}. The spin polarization vector is incorporated into the algorithm as an additional quantity and calculated for each free carrier~\cite{Spezia2010_2}. All simulations are performed in a GaAs crystal with a doping concentration $n$ equal to $10^{16}$ cm$^{-3}$ and a lattice temperature $T_L$=$300$ K. Moreover, we assume that all donors are ionised and that the free electron concentration is equal to the doping concentration. An ensemble of $5\cdot10^{4}$ electrons is used to collect spin statistics.  All physical quantities of interest are calculated after a transient time long enough to achieve the steady-state transport regime. The spin relaxation simulation starts with all electrons in the $\Gamma$ valley and initially polarized $(\vect{S}(0) = {\bf 1})$ along the $\hat{{\bf x}}$-axis of the crystal, at the injection plane $(x_{0} = 0)$. The spin relaxation time $\tau$ is calculated by extracting the time corresponding to a reduction of the initial spin polarisation by a factor $1/e$. Our model has been validated by the experimental results reported in Ref.~\cite{Viana2012} and also provides spin lifetimes in good agreement with those calculated in the recent theoretical work of Tong and Wu ~\cite{Tong2012}. For statistical purposes, in the presence of an external source of noise, we performed $500$ different realizations, and evaluated both average values and error bars of the all extracted physical quantities.

\section{Numerical results and discussion}

\indent In panel (a) of fig.~\ref{fig.1}, we show the \textit{normalized} electron spin relaxation time, that is the ratio between $\tau$ and $\tau_0$, as a function of the \textit{normalized} correlation time $\tau_D/\tau_0$, for different values of the noise amplitude $\Delta$. Here, $\tau_0= 1.40$ ps is the value of the electron spin lifetime numerically obtained in the absence of external noise, $\tau$ is the electron spin relaxation time as modified by the DM noise. The deterministic field $F_0 = 6$ kV/cm is greater than the Gunn field $F_G\approx3.25$ kV/cm, i.e. the minimum value of electric field that the electrons need to move in L-valleys. 
\begin{figure*}
\includegraphics[scale=0.95]{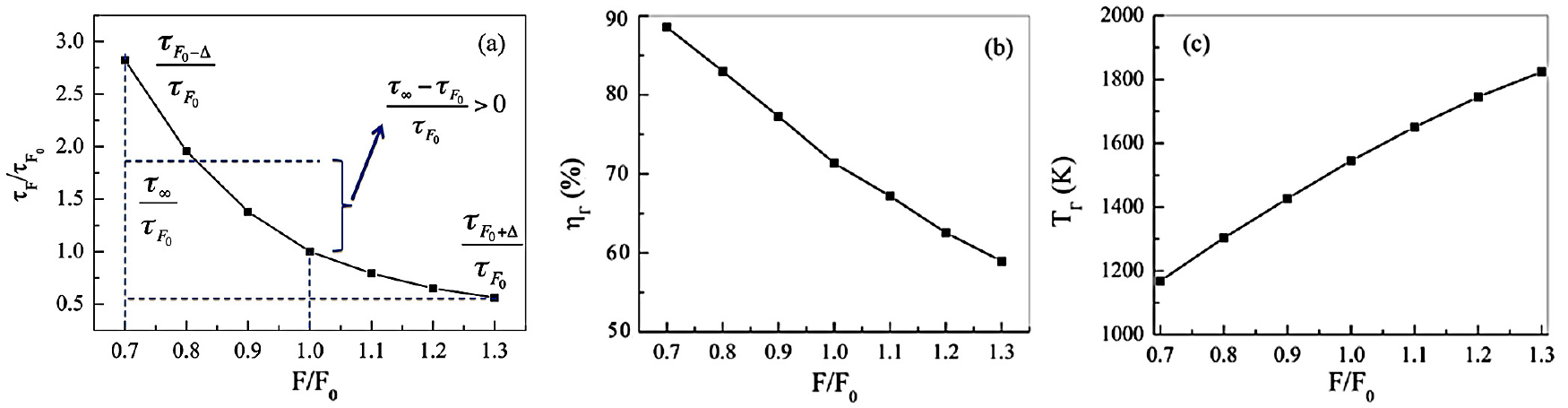}
\caption{(a) Normalized electron spin relaxation time $\tau_F/\tau_{F_0}$, (b) electron occupation percentages and (c) hot-electron temperature in $\Gamma$-valley, as a function of the ratio $F/F_0$. $\tau_{F_0-\Delta}$, $\tau_{F_0+\Delta}$ are the spin relaxation times obtained with 
$F=F_0-\Delta$ and $F=F_0+\Delta$, with different values of noise amplitude and static electric field: $F_0=6$ kV/cm, $\Delta = 0.6, 1.2, 1.8$ kV/cm. Here, 
$\tau_\infty=(\tau_{F_0-\Delta}+\tau_{F_0+\Delta})/2$. The values of the other parameters are: $n=10^{16}$ cm$^{-3}$, $T_L$=$300$ K.}
\label{fig.2}
\end{figure*}
For the lowest noise amplitude $\Delta = 0.6$ kV/cm, the electron spin relaxation time is almost constant ($\tau/\tau_0 \sim 1$). For values of the noise correlation time $\tau_D \le 10^{-1}\tau_0$, the value of $\tau$ is always close to $\tau_0$, even at higher values of $\Delta$. On the contrary, when the noise amplitude increases and $\tau_D > \tau_0$, the value of the spin relaxation time $\tau$ can increase up to $1.25~\tau_0$. For 
$\tau_D > 10~\tau_0$, the electron spin lifetime remains approximately constant. This positive effect monotonically increases with the amplitude of the DM noise. Panels (b) and (c) of fig.~\ref{fig.1} show the effect of the addition of a DM noise component to the driving electric field on the electron occupation percentage and the hot-electron temperature in $\Gamma$-valley, respectively. Our findings show that, in the presence of DM noise with $\tau_D >~10\tau_0$, a greater number of electrons remains in the $\Gamma$-valley and the hot-electron temperature in $\Gamma$-valley is slightly reduced. We note that, due to the nonlinearity of the spin relaxation process, little variations of $\eta_\Gamma$ and  $T_\Gamma$, induced by the addition of noise, can be responsible of the observed increase of the spin relaxation lifetimes. This is a very interesting point that deserve further investigations. The benefits of the dichotomous noise are grounded on this threshold effect, in which an enhancement of the electron spin lifetime can be maintened for several orders of magnitude of the DM mean switching time, starting from a value equal to $10$ times the relaxation characteristic time of the spin system in absence of noise, i.e. the spin lifetime $\tau_0$.\\ 
\indent By a simple argument it is possible to explain the numerical results found in the presence of an external source of DM noise. In the limit $\tau_D \rightarrow 0$, the spin dynamics is characterized by a high frequency switching rate between the two states of electric field. Therefore, being the spin relaxation time much greater than $\tau_D$, each electron of the electron spin ensemble will experience a mean electric field, which coincides with the deterministic component $F_0$. As a consequence, $\tau$ slightly deviates from $\tau_0=\tau_{F_0}$ (see fig.~\ref{fig.1}a). In the opposite limit $\tau_D\rightarrow\infty$, we have to distinguish between two cases depending on the initial value of the electric field, namely $F_0-\Delta$ or $F_0+\Delta$. In the first case the spin relaxation time is $\tau_{F_0-\Delta}$ corrected, in a first approximation, by the spin relaxation time $\tau_{F_0+\Delta}$, if the dichotomous source switches, in average, into the state $F_0+\Delta$ before the relaxation time $\tau_{F_0-\Delta}$. This occurs with a probability $p_L=\frac{1}{2}\left[1-\exp{\left(-\tau_{F_0-\Delta}/\tau_D\right)}\right]$. This first-order correction gives a contribution of $ \tau_- = \tau_{F_0-\Delta}(1-p_L)+\tau_{F_0+\Delta}(p_L)$ to the total depolarization time. Likewise, in the second case the additional contribution is $\tau_+ = \tau_{F_0+\Delta}(1-p_H)+\tau_{F_0-\Delta}(p_H)$ and the occurrence probability is $p_H=\frac{1}{2}\left[1-\exp{\left(-\tau_{F_0+\Delta}/\tau_D\right)}\right]$. Therefore, the total spin relaxation time is obtained as the arithmetic average of the previous contributions\\
\begin{eqnarray}
\label{tau1}
\tau &=& \frac{\tau_- + \tau_+}{2} = \frac{\tau_{F_0-\Delta}+\tau_{F_0+\Delta}}{2}+\frac{\tau_{F_0-\Delta}-\tau_{F_0+\Delta}}{4} \nonumber\\
&&\cdot\left[\exp{\left(-\tau_{F_0-\Delta}/\tau_D\right)}
-\exp{\left(-\tau_{F_0+\Delta}/\tau_D\right)}\right].
\end{eqnarray}
\indent  By expanding the exponential function as $e^x \approx 1 + x$, in the limit $\tau_D\rightarrow\infty$, we get
\begin{equation}
\label{tau2}
%\tau\approx\frac{\tau_{F_0-\Delta}+\tau_{F_0+\Delta}}{2}-\frac{\left(\tau_{F_0-\Delta}-\tau_{F_0+\Delta}\right)^2}{4\tau_D},
\tau_\infty=(\tau_{F_0-\Delta}+\tau_{F_0+\Delta})/2.
\end{equation}
\begin{figure*}
\includegraphics[width=5.8cm]{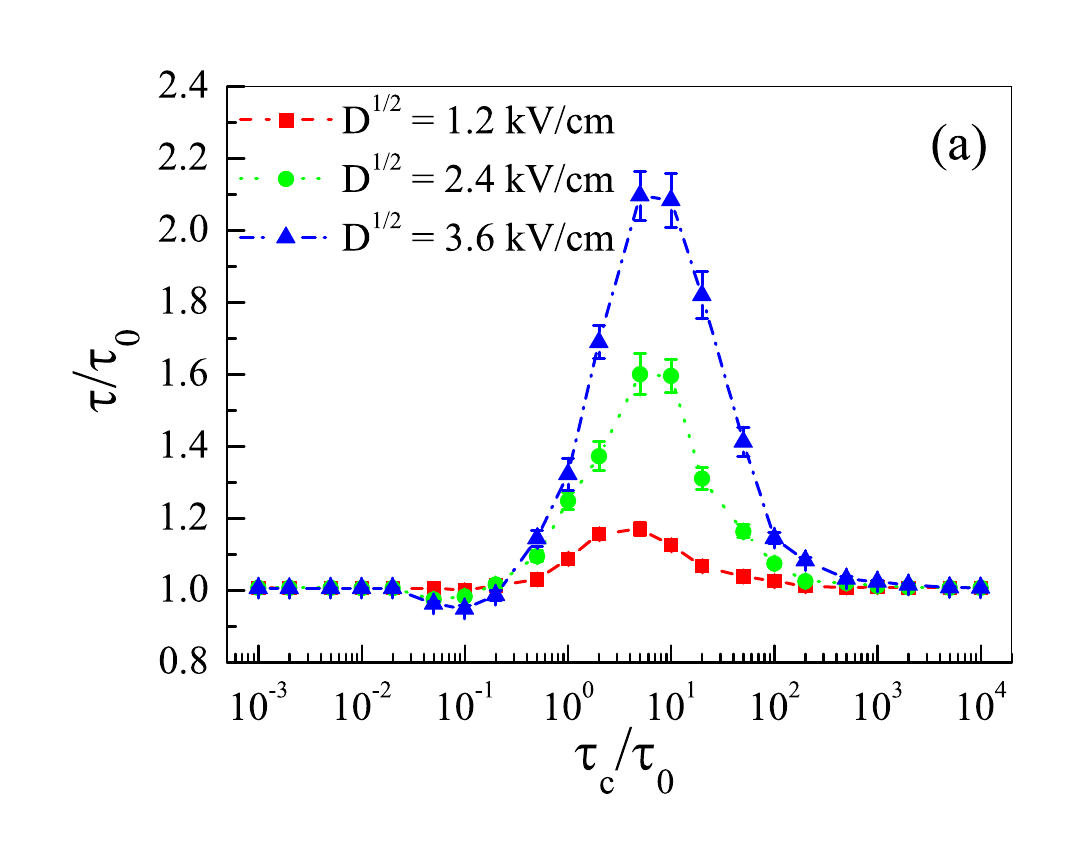}
\includegraphics[width=5.8cm]{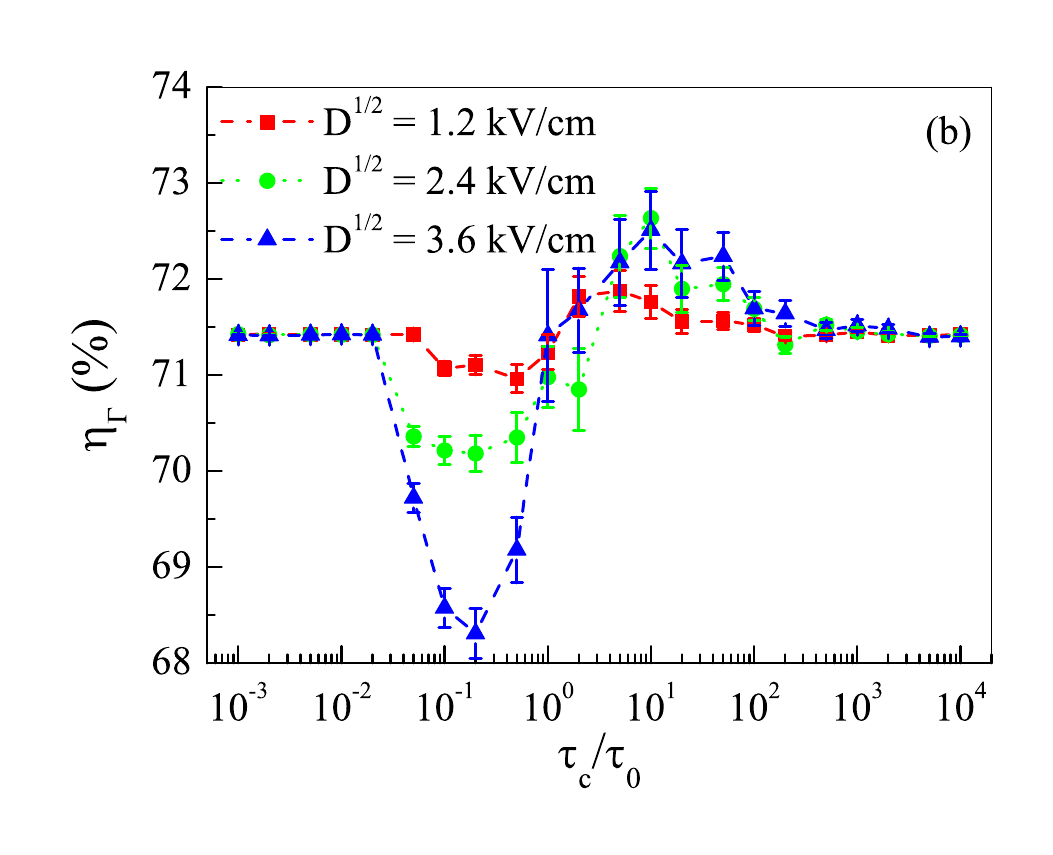}
\includegraphics[width=5.8cm]{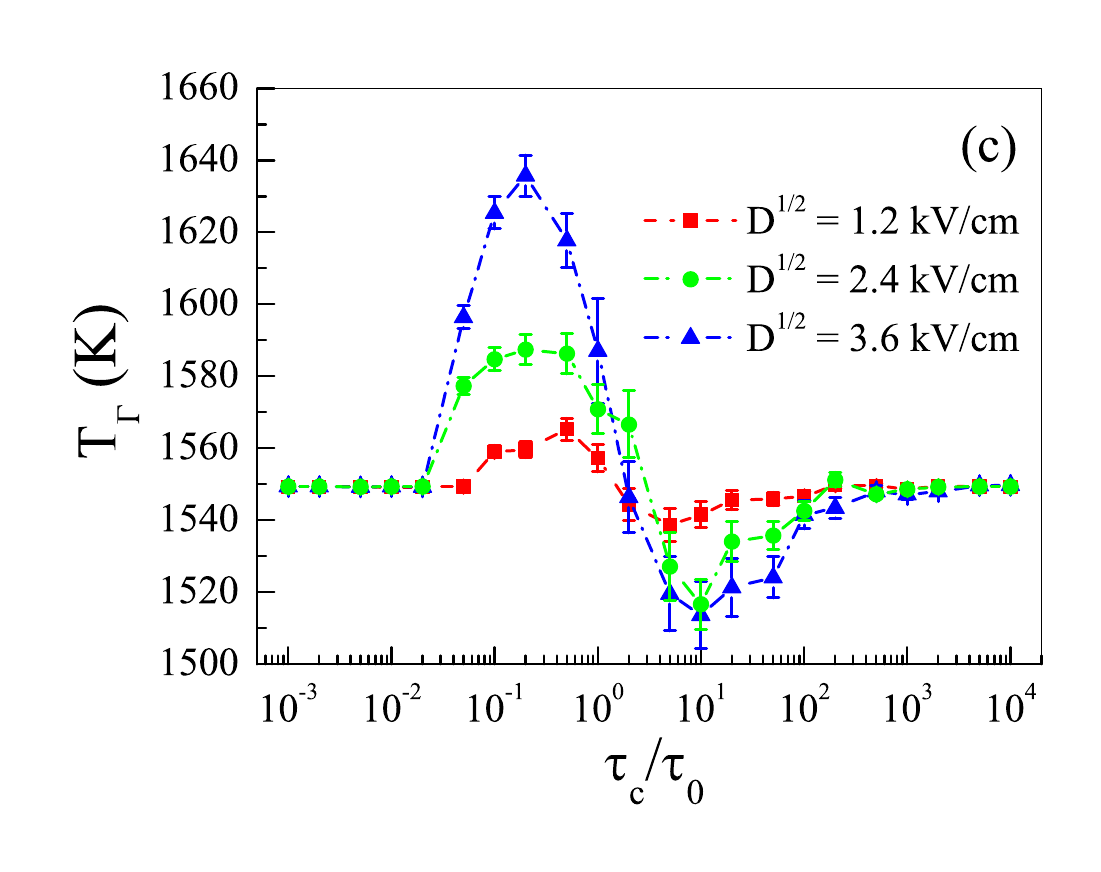}
\caption{Gaussian correlated noise results: (a) Normalized electron spin relaxation time $\tau/\tau_0$, (b) electron occupation percentage $\eta_\Gamma$ and (c) hot-electron temperature $T_\Gamma$ in $\Gamma$-valley, as a function of the normalized noise correlation time $\tau_c/\tau_0$, for three different values of noise amplitude, namely $D^{1/2} = 1.2, 2.4, 3.6$ kV/cm. The values of the other parameters are: $n=10^{16}$ cm$^{-3}$, $T_L$=$300$ K, $F_0=6$ kV/cm and $\tau_0=1.40$ ps.}
\label{fig.3}
\end{figure*}
To analyze in more detail the behaviour of $\tau$ as a function of the correlation time $\tau_D$ we have calculated the normalized electron spin relaxation time, that is the ratio between  $\tau_F$ and $\tau_{F_0}$ obtained with the constant electric fields $F$ and $F_0$, respectively, as a function of the ratio $F/F_0$. The results are shown in panel (a) of fig.~\ref{fig.2}, where we show the values of $\tau_{F_0-\Delta}$  and $\tau_{F_0+\Delta}$, obtained with constant amplitude of the applied electric field $F_0-\Delta$  and $F_0+\Delta$, respectively. In the same panel we also indicate the asymptotic value $\tau_\infty$, obtained for example with $\Delta=30\%$ of $F_0$. We obtain $\tau_\infty/\tau_{F_0}\approx 1.75$, which is a good estimation of the asymptotic value of $\tau$ shown in fig.~\ref{fig.1}a. The shape of the curve of $\tau_F/\tau_{F_0}$, which is a concave function of $F/F_0$, confirms that the difference $\tau_\infty-\tau_{F_0}$ is always greater than zero.  With similar considerations, by using panels (b) and (c) of fig.~\ref{fig.2}, we can take into account the little variations of the occupation percentage and the hot-electron temperature in $\Gamma$-valley, shown in fig.~\ref{fig.1}  b and c.\\
\indent In the case of the fluctuating driving electric field, due to the presence of  an external source of GC noise, the dependence of the normalized spin relaxation time $\tau/\tau_0$ on the normalized noise correlation time $\tau_c/\tau_0$, is shown in fig.~\ref{fig.3}a, for three different values of the noise intensity. A detailed analysis of these data highlights the presence of a nonmonotonic behaviour characterized by a slight minimum at $\tau_c/\tau_0\sim10^{-1}$ and a more evident maximum for a value of noise correlation time of about 10 $\tau_0$. Our numerical results show that the addition of a Gaussian correlated noise, with  $\tau_c$ in the range ($1 \div 100) \tau_0$, causes an enhancement of the value of the spin relaxation time $\tau$ which may increase up to $\sim 2.1$ $\tau_0$ depending on the value of noise intensity $D$. For very low and very high values of $\tau_c$, the time $\tau$ remains close to $\tau_0$. These results are in agreement with the nonmonotonic behaviour of the spin depolarization length versus the normalized noise correlation time, obtained in a previous investigation but with lower values of free electron concentration and temperature ($n=10^{13}$ cm$^{-3}$ and $T_L$=$77$ K)~\cite{Spezia2012}.
Panels (b) and (c) of fig.~\ref{fig.3} show the effect of the addition of a GC noise component to the driving electric field on the electron occupation percentage and the hot-electron temperature in $\Gamma$-valley, respectively. We find that, in the presence of the GC noise, the electron occupation percentage $\eta_\Gamma$ shows a non monotonic behaviour, with a minimum at $\tau_c/\tau_0\sim10^{-1}$, and a slight maximum at about 10 $\tau_0$ for all the different values of noise intensity $D$. The hot-electron temperature shows an opposite non monotonic behaviour, characterized by an increase at $\tau_c/\tau_0\sim10^{-1}$, and a reduction at about 10 $\tau_0$.\\ 
The complex behaviour observed in the spin lifetimes can be ascribed to the effects produced by the  characteristics of the added external noise, structurally different from the dichotomous fluctuations, and it cannot be explained by using a simple argument, as in the DM case. In fact, in the presence of GC noise the electron ensemble experiences an effective electric field that can be higher or lower than the deterministic one, depending on the value of the noise correlation time, which represents the characteristic memory time of the fluctuations. This affects the electron transport in the semiconductor in a way that an enhancement or a reduction of the electron spin lifetime, occupation percentage and hot-electron temperature can be obtained. The occurrence of these circumstances depends not only on the ratio between the value of the memory time of the GC noise and the characteristic relaxation time of the spin system, but also on its ratio with both the momentum relaxation time and the momentum redistribution time (characteristic of the electron-electron interaction). Thus, if the GC noise becomes "resonant" with one of these characteristic times of the systems, it can give rise to a constructive or destructive interference, and produce significant changes on the electron dynamics.

\section{Conclusions}
In this letter we have investigated the influence of two different sources of fluctuations on the electron spin relaxation process in $n$-doped GaAs semiconductor crystals, mainly focusing on the role of a symmetric dichotomous noise. Our findings show that the addition of a source of DM noise to a static driving electric field, whose amplitude is greater than the Gunn threshold, can enhance the spin lifetime up to 25\% of its value in the absence of external noise. This enhancement, which increases with the amplitude of the external fluctuations, is observed for noise correlation times comparable to or greater than the spin lifetime obtained without added fluctuations. 
The positive effect, ascribed to the different effective electric field experienced by the electron ensemble, within the time window of the spin relaxation time, is associated to a decrease of the occupation of the $L$-valleys, where the strength of spin-orbit coupling felt by electrons is at least one order of magnitude greater than that present in $\Gamma$-valley.  \\
\indent In the case of Gaussian correlated noise, a nonomonotonic behaviour of the spin relaxation time is observed, with a maximum for values of the noise correlation time close to $10$ times the spin dephasing time, obtained in the absence of added noise. In this case, depending on the value of the noise correlation time, the electron ensemble experiences an effective electric field that can be higher or lower than the deterministic one. In particular, when the noise memory time is comparable with one of the characteristic times of the electron dynamics, a constructive or distructive interference effect occurs and gives rise to an enhancement or a reduction of the spin relaxation time.
\\
\indent In summary, the possible enhancement of the electron spin lifetime in GaAs crystals strongly depends on the correlation time and amplitude of the external noise. In particular, we have found that the benefits of the dichotomous noise consist of a threshold effect, in which the increase of the electron spin lifetime is obtained in a wide range of noise correlation times, while in the presence of a gaussian correlated component, the enhancement is greater, but is obtainable only in a limited range of correlation times. In conclusion, random fluctuations of the electric driving field, due to different noise sources, can play a relevant role on controlling and tuning the coherence of spin-relaxation processes. In this view, by using appropriate noise characteristic times, it is possible to select the most favorable condition for the transmission of information by electron spin. 

\acknowledgments
This work was partially supported by CNISM and by MIUR through Grant. No. PON02 00355 3391233, ENERGETIC.

\end{document}